\documentclass[12pt]{amsart}
\usepackage{amsfonts}
\usepackage{amssymb,latexsym}
\theoremstyle{plain}
\numberwithin{equation}{section}
\newtheorem{theorem}{Theorem}
\newtheorem{proposition}{Proposition}
\newtheorem{lemma}{Lemma}
\newtheorem{corollary}{Corollary}

\newtheorem{remark}{Remark}
\newtheorem{definition}{Definition}
\begin{document}

%AccBou(Heisenberg-CE5).tex

\title[Central extensions of the Heisenberg algebra]{Central extensions of the Heisenberg algebra}

\author{Luigi Accardi}
\address{Centro Vito Volterra, Universit\`{a} di Roma Tor Vergata\\
            via Columbia  2, 00133 Roma, Italy}
\email{accardi@volterra.mat.uniroma2.it}
\urladdr{http://volterra.mat.uniroma2.it}
\author{Andreas Boukas}
\address{Department of Mathematics and Natural Sciences, American College of Greece\\
 Aghia Paraskevi, Athens 15342, Greece}
\email{andreasboukas@acgmail.gr}

\date{\today}

\begin{abstract}
We study the non-trivial central extensions $CEHeis$ of the Heisenberg algebra $Heis$ recently constructed in \cite{AccBouCE}. We prove that a real form of $CEHeis$ is one the fifteen classified real four--dimensional solvable Lie algebras. We also show that $CEHeis$ can be realized (i) as a sub--Lie--algebra of the Schroedinger algebra and (ii) in terms of two independent copies of the canonical commutation relations (CCR).  This gives a natural family of unitary representations of $CEHeis$ and allows an explicit determination of the associated group by exponentiation. In contrast with $Heis$, the group law for $ CEHeis$ is given by nonlinear (quadratic) functions of the coordinates.
\end{abstract}

\maketitle

\section{Central extensions of the Heisenberg algebra}

\subsection{The $CEHeis$ $*$--Lie algebra.}  The generators $a$, $a^{\dagger}$ and $h$ of the (one mode) Heisenberg algebra $Heis$ satisfy the Lie algebra commutation relations

\[
\lbrack a,a^{\dagger}\rbrack_{Heis}=h\qquad;\qquad\lbrack a, h\rbrack_{Heis }=\lbrack  h,  a^{\dagger}\rbrack_{Heis }=0\label{1.1}
\]

\noindent and the duality relations

\begin{equation}\label{dualrel}
(a)^*=a^{\dagger}\,\,;\,\,h^*=h
\end{equation}

\medskip

\noindent As shown in \cite{AccBouCE} this algebra admits non trivial central extensions. More precisely, all $2$-cocycles $\phi$ on $Heis \times  Heis$ are defined through a bilinear skew-symmetric extension of the functionals

\begin{equation}\label{1.3}
\phi( a, a^{\dagger})=\lambda
\qquad;\qquad\phi ( h, a^{\dagger})= z 
\qquad;\qquad\phi ( a ,  h)= \bar z 
\end{equation}

\noindent where $\lambda \in \mathbb{R}$ and $z \in \mathbb{C}$. 
Each $2$-cocycle (\ref{1.3}) defines a central  extension 
$ CEHeis$ of $Heis$ and this central extension  is trivial if and only if $z=0$. Duality relations (\ref{dualrel}) still hold.

\medskip

\noindent The centrally extended Heisenberg commutation relations (CECCR) are

\begin{equation}\label{extcomm}
\lbrack  a, a^{\dagger} \rbrack_{CEHeis}= 
h+\lambda \, E \,\,;\,\,
\lbrack  h, a^{\dagger} \rbrack_{CEHeis}= z \, E \,\,;\,\,
\lbrack a , h  \rbrack_{CEHeis}=\bar z \, E 
\end{equation}

\noindent where $E\not\equiv 0$ is the self-adjoint central element and where, here and in the following, all omitted commutators are assumed to be equal to zero.

\medskip

\noindent Renaming $h+\lambda\,E$ by just $h$ in (\ref{extcomm}) we obtain the equivalent commutation relations

\begin{equation}\label{extcommeq}
\lbrack  a, a^{\dagger} \rbrack_{CEHeis}= h \qquad;\qquad
\lbrack  h, a^{\dagger} \rbrack_{CEHeis}= z \, E \qquad;\qquad
\lbrack a , h  \rbrack_{CEHeis}=\bar z \, E 
\end{equation}

\noindent From now on we will use (\ref{extcommeq}) and (\ref{dualrel}) as the defining commutation relations of $CEHeis$.  For $z=0$ we are back in the usual Heisenberg algebra. In the following we assume that $z\not=0$.

\begin{proposition} 
For $\lambda \in \mathbb{R}$ and $z \in \mathbb{C}-\{0\}$ commutation relations (\ref{extcommeq}) define a solvable four--dimensional $*$--Lie algebra $CEHeis$ with generators $a, a^{\dagger}, h$ and $E$.
\end{proposition}

\begin{proof} Let $l_1=a,l_2=a^{\dagger},l_3=h,l_4=E$. Using (\ref{extcommeq}) we have that

\[
\lbrack l_2, l_3 \rbrack_{CEHeis}=-zE\,\,;\,\,\lbrack l_3, l_1 \rbrack_{CEHeis}=-\overline zE\,\,;\,\,\lbrack l_1, l_2 \rbrack_{CEHeis}=h
\]

\noindent Hence

\[
\lbrack  l_1 , \lbrack l_2, l_3 \rbrack_{CEHeis} \rbrack_{CEHeis}=\lbrack  l_2 , \lbrack l_3, l_1 \rbrack_{CEHeis} \rbrack_{CEHeis}=\lbrack  l_3 , \lbrack l_1, l_2 \rbrack_{CEHeis} \rbrack_{CEHeis}=0
\]

\noindent which implies that

\[
\lbrack  l_1 , \lbrack l_2, l_3 \rbrack_{CEHeis} \rbrack_{CEHeis}+\lbrack  l_2 , \lbrack l_3, l_1 \rbrack_{CEHeis} \rbrack_{CEHeis}+\lbrack  l_3 , \lbrack l_1, l_2 \rbrack_{CEHeis} \rbrack_{CEHeis}=0
\]

\medskip

\noindent i.e.  the Jacobi identity is satisfied.  To show that $a, a^{\dagger}, h$ and $E$ are linearly independent, suppose that

\begin{equation}\label{li}
\alpha \,a+\beta \,a^{\dagger}+\gamma \,h+\delta \,E=0
\end{equation}

\noindent where $\alpha, \beta, \gamma, \delta\in\mathbb{C}$. Taking the commutator of (\ref{li}) with $a^{\dagger}$ we find that

\[
\alpha \,h+\gamma z E=0
\]

\noindent which, after taking its commutator with $a^{\dagger}$, implies that $\alpha \,\,z \,E=0$. Since $z\neq0$ and $E\not\equiv0$, it follows that  $\alpha=0$ and (\ref{li}) is reduced to

\begin{equation}\label{li2}
\beta \,a^{\dagger}+\gamma \,h+\delta \,E=0
\end{equation}

\noindent Taking the commutator of (\ref{li2}) with $h$ we find that $\beta \,z \,E=0$. Hence $\beta=0$ and (\ref{li2}) is reduced to

\begin{equation}\label{li3}
\gamma \,h+\delta \,E=0
\end{equation}

\noindent Taking the commutator of (\ref{li3}) with $a^{\dagger}$ we find that $\gamma \,z \,E=0$. Hence $\gamma=0$ and (\ref{li3}) is reduced to

\[
\delta \,E=0
\]

\noindent which implies that $\delta=0$ as well.  Finally 

\[
{CEHeis}^{(1)}:=\lbrack CEHeis,  CEHeis \rbrack
=\{\gamma \,h+\delta \,E\,\,\, : \,\,\,  \gamma, \delta\in\mathbb{C} \}
\]

\noindent and

\[
{CEHeis}^{(2)}:=
\lbrack CEHeis^{(1)}, CEHeis^{(1)}\rbrack =\{0\}
\]

\noindent Therefore $CEHeis$ is solvable.
\end{proof}

\subsection{Real form of $CEHeis$.} For an operator $X$, in particular for a complex number, we denote by $Re \, X$ and $Im \,X$ its real and imaginary part defined respectively by

\[
Re \, X := \frac{1}{2}(X+X^*) \qquad ; \qquad   Im \, X:= \frac{1}{2i}(X-X^*) 
\]

\noindent By construction, $X=Re \, X + i\,Im \,X$ and the right hand side is called the {real decomposition of} $X$. 

\begin{proposition}\label{real}  
In the above notations define $p$, $q$ and $H$ by

\begin{equation}\label{dfpqH}
a^{\dagger}=p+i\,q  \qquad ; \qquad  a=p-i\,q  \qquad ; \qquad   H= -ih/2
\end{equation}

\noindent  Then $p, q, E$ are self-adjoint and $H$ is skew-adjoint. Moreover $p, q, E$ and $H$ are the generators of a real four-dimensional solvable $*$--Lie algebra with central element $E$ and commutation relations

\begin{equation}\label{rextcomm}
\lbrack  p, q \rbrack= H\,\,;\,\,\lbrack  q, H\rbrack= c \, E \,\,;\,\,
\lbrack H , p  \rbrack=b\, E 
\end{equation}

\noindent  where $b, c$ are (not simultaneously zero) real numbers given by

\begin{equation}\label{rels-bcz}
c=\frac{Re \,z}{2} \ , \ b=\frac{Im\, z}{2}
\end{equation}

\noindent  Conversely, let $p, q, H, E$ be the generators (with $p, q, E$ self-adjoint and $H$ skew-adjoint) of a real four-dimensional solvable $*$--Lie algebra with central element $E$ and commutation relations (\ref{rextcomm}) where $b$ and $c$ are (not simultaneously zero) real numbers. Then, defining $z$ by (\ref{rels-bcz}), the operators defined by (\ref{dfpqH}) are the generators of the nontrivial central extension $CEHeis$ of the Heisenberg algebra defined by (\ref{extcommeq}), (\ref{dualrel}).

\end{proposition}

\begin{proof} The proof consists of a simple algebraic verification.
\end{proof}

\subsection{Matching of the real form of $CEHeis$ with the real four--dimensional solvable Lie algebras classification list.} Real four--dimensional solvable Lie algebras are fully classified. There are exactly fifteen isomorphism classes and they are listed, for example, in proposition 2.1 of \cite{Ovando} (see references therein for additional information). One of the fifteen Lie algebras that appear in the above mentioned classification list is the Lie algebra  denoted by $\eta_4$ with generators $e_1, e_2, e_3, e_4$ and (non-zero) commutation relations

\begin{equation}\label{n4}
\lbrack e_4, e_1 \rbrack  = e_2\qquad ; \qquad  \lbrack e_4, e_2 \rbrack = e_3
\end{equation}

\begin{corollary} 
The real four--dimensional solvable Lie algebra described in proposition \ref{real} can be identified to the algebra $\eta_4$  defined above.
\end{corollary}

\begin{proof} 
 \noindent In the notation (\ref{rels-bcz}) if $b=0$ and $c\neq0$ then we may take 

\[
e_4=q\quad ; \quad  e_1=p \quad ; \quad   e_2=-H \quad ; \quad  e_3=-c\,E
\]
 
\noindent  If $c=0$ and $b\neq0$ then we may take 

\[
e_4=p\quad ; \quad e_1=q\quad ; \quad e_2=H\quad ; \quad e_3=-b\,E
\]
 
\noindent  If both $b$ and $c$ are not equal to zero then replacing $q$ by $\hat q=\alpha \, p+\beta\,q$ in (\ref{rextcomm}), where $ \alpha , \beta \in\mathbb{R}\setminus \{0\}$ are such that $\beta c-\alpha b=0$, we obtain the commutation relations

\[
\lbrack p, \hat q \rbrack =\beta \, H\,\,\,;\,\,\,\lbrack  \hat q , H \rbrack=0\,\,\,;\,\,\, \lbrack H, p \rbrack =b\,E
\] 

\noindent which, letting $\hat p:= \frac{p}{\beta}$ and $d:= \frac{b}{\beta}$, become

\[
\lbrack \hat p, \hat q \rbrack = H\,\,\,;\,\,\,\lbrack  \hat q , H \rbrack=0\,\,\,;\,\,\, \lbrack H, \hat p \rbrack = d\,E
\] 

\noindent Denoting 

\[
e_4=\hat p\quad ; \quad e_1=\hat q\quad ; \quad e_2=H \quad ; \quad e_3= -d\,E
\]

\noindent we obtain the commutation relations (\ref{n4}).
\end{proof}

\section{Representations of  $CEHeis$} 

\subsection{Realization of $CEHeis$ as a proper sub--algebra of the Schroedinger algebra} In this subsection we show how the generators $a$, $a^{\dagger}, h$ and $E$ of $CEHeis$ can be expessed in terms of the generators of the Schroedinger algebra.

\begin{definition}\label{sch} 
The Schroedinger algebra is the six--dimensional $*$--Lie algebra generated by $b$, $b^{\dagger}$, $b^2$, ${b^{\dagger}}^2$, $b^{\dagger}\,b$ and $1$ where $b^{\dagger}$, $b$ and $1$ are the generators of a Boson Heisenberg algebra with

\begin{equation}
\lbrack b, b^{\dagger} \rbrack=1\qquad;\qquad{(b^{\dagger})}^*=b\label{ccrb}
\end{equation}

\end{definition}

\begin{lemma}\label{sf} In the notation of definition \ref{sch}

\noindent (i) $\lbrack    b-b^{\dagger}, b+ b^{\dagger} \rbrack=2$

\medskip

\noindent (ii) $\lbrack (b-b^{\dagger})^2, b+b^{\dagger} \rbrack=  4\,(b-b^{\dagger})$, where $(b-b^{\dagger})^2=b^2+{b^{\dagger}}^2-2\,b^{\dagger}\,b-1$

\medskip

\noindent (iii) For an analytic function $f$, $b\,f(b^{\dagger})=f(b^{\dagger})\,b+f^{\prime}(b^{\dagger})$

\end{lemma}

\begin{proof} The proof of (i) and (ii) is straight--forward. Part  (iii) is proposition 2.1.1 of \cite{Fein}.
\end{proof}

\begin{theorem}\label{Schr}(Boson representation of $CEHeis$) 
 Let $\lbrack b, b^{\dagger} \rbrack=1$ with ${(b^{\dagger})}^*=b$. 

\medskip

\noindent (i) If $z\in\mathbb{C}$ with $Re\, z\neq0$, then for arbitrary $\rho, r\in\mathbb{R}$ with $r\neq0$, define:

\begin{eqnarray}
a&:=&\left( \frac{4\,\rho\,Im \,z-r^2 }{4\,Re \,z}+i\,\rho \right)\,( b- b^{\dagger}  )^2
-\frac{ i\,\bar z}{2\,r }\,(  b+ b^{\dagger} ) \label{dfarznn}\\
a^{\dagger}&:=&\left( \frac{4\,\rho\,Im \,z-r^2 }{4\,Re \,z}-i\,\rho \right)\,
( b- b^{\dagger}  )^2+\frac{ i\, z}{2\,r }\,(  b+ b^{\dagger} ) \label{dfa+rznn}
\end{eqnarray}

\noindent and

\begin{equation}
h:=i\,r\,(  b^{\dagger}-b)\label{dfhrznn}
\end{equation}

\noindent The quadruple $\{a^+, a, h, E=1\}$ satisfies the  commutation relations (\ref{extcommeq}) and the duality relations (\ref{1.1}) of $CEHeis$.

\medskip

\noindent (ii) If $z\in\mathbb{C}$ with $Re \,z=0$,  then 
for arbitrary $\rho, r\in\mathbb{R}$ with $r\neq0$, define

\begin{eqnarray}
a&:=&\left(\rho-\frac{i\,Im \,z }{16\,r^2} \right)\,( b- b^{\dagger}  )^2+r\,
(  b+ b^{\dagger} ) \label{dfarzn}\\
a^{\dagger}&:=&\left(\rho+\frac{i\,Im \,z }{16\,r^2} \right)\,( b- b^{\dagger}  )^2+r\,
(  b+ b^{\dagger} )  \label{dfa+rzn}
\end{eqnarray}

\noindent and 

\begin{equation}
h:=\frac{i\,Im \,z }{2\,r} \,(b^{\dagger}-b)\label{dfhrzn}
\end{equation}

\noindent The quadruple $\{a^+, a, h, E=1\}$ satisfies the commutation relations (\ref{extcommeq}) and the duality relations (\ref{1.1}) of $CEHeis$.
\end{theorem} 

\begin{proof} To prove (i), using lemma \ref{sf} we have

\begin{eqnarray*}
&\lbrack  a, a^{\dagger} \rbrack=&\\
& \left( \frac{4\,\rho\,Im \,z -r^2}{4\,Re \,z}+i\,\rho \right)\,\frac{ i\, z}{2\,r }\,\lbrack (b-b^{\dagger})^2, b+b^{\dagger} \rbrack-\frac{ i\,\bar z}{2\,r }  \,\left( \frac{4\,\rho\,Im \,z -r^2}{4\,Re \,z}- i\,\rho \right)\,\lbrack  b+ b^{\dagger},  (b-b^{\dagger})^2 \rbrack=&\\
&\left( \frac{4\,\rho\,Im\, z -r^2}{4\,Re\, z}+i\,\rho \right)\,\frac{ i\, z}{2\,r }\,4\,(b-b^{\dagger})    -\frac{ i\,\bar z}{2\,r }  \,\left( \frac{4\,\rho\,Im \,z -r^2}{4\,Re \,z}- i\,\rho \right)\,4\,(b^{\dagger}-b) =-i\,r\,( b- b^{\dagger})=h&
\end{eqnarray*}

\medskip

\noindent Similarly

\[
\lbrack  a, h \rbrack=-\frac{ i\,\bar z}{2\,r }  \,(-i\,r)\,\lbrack    b+b^{\dagger}, b- b^{\dagger} \rbrack=-\frac{ \bar z\,r}{2\,r }  \,(-2)=\bar z
\]

\noindent and

\[
\lbrack  h, a^{\dagger} \rbrack=(-i\,r)\,\frac{ i\,z}{2\,r }  \,\lbrack    b-b^{\dagger}, b+b^{\dagger} \rbrack=\frac{  z\,r}{2\,r }  \,2= z
\]

\noindent Clearly $(a^{\dagger} )^*=a$ and $h^*=h$. The proof of (ii) is similar.

\end{proof}

\begin{definition}\label{Fock} For $\lambda\in\mathbb{C}$ let $y(\lambda)=e^{\lambda\,b}$. The Heisenberg Fock space $\mathcal{F}$ is the Hilbert space completion of the linear span of the exponential vectors $\{y(\lambda)\,;\,\lambda\in\mathbb{C}\}$ with respect to the inner product

\begin{equation}
\langle y(\lambda), y(\mu) \rangle=e^{\bar \lambda\,\mu}
\end{equation}

\end{definition}

\noindent It is well known that 

\begin{equation}\label{b1}
b\,y(\lambda)=\lambda\,y(\lambda)
\end{equation}

\noindent and

\begin{equation}
b^{\dagger}\,y(\lambda)=\frac{\partial}{\partial\,\epsilon}|_{\epsilon=0}\,y(\lambda+\epsilon)
\end{equation}

\noindent Therefore, for non-negative integers $n, k$ 

\begin{eqnarray}
b^k\,y(\lambda)&=&\lambda^k\,y(\lambda)\\
{b^{\dagger}}^n\,y(\lambda)&=&\frac{\partial^n}{\partial\,\epsilon^n}|_{\epsilon=0}\,y(\lambda+\epsilon)
\end{eqnarray}

\noindent and in general

\begin{equation}\label{b2}
{b^{\dagger}}^n\,b^k\,y(\lambda)=\lambda^k\,\frac{\partial^n}{\partial\,\epsilon^n}|_{\epsilon=0}\,y(\lambda+\epsilon)
\end{equation}

\begin{theorem}\label{F}(Boson Fock representation of $CEHeis$) In the notation of theorem \ref{Schr} and definition \ref{Fock}:

\medskip
 
\noindent (i) If $z\in\mathbb{C}$ with $Re z\neq0$ then

\begin{eqnarray}
&a\,y(\lambda)=\left( \left(\frac{4\,\rho\,Im \,z-r^2 }{4\,Re\, z}+i\,\rho \right)\,( \lambda^2-1)- \frac{ i\,\bar z}{2\,r }\,\lambda \right)  \,y(\lambda)&\\
&+\left( \left(\frac{4\,\rho\,Im \,z-r^2 }{4\,Re \,z}+i\,\rho \right)\,\frac{\partial^2}{\partial\,\epsilon^2}|_{\epsilon=0}-\left( \left(\frac{4\,\rho\,Im \,z-r^2 }{4\,Re \,z}+i\,\rho \right)\,2\,\lambda+\frac{ i\,\bar z}{2\,r }\right)\,\frac{\partial}{\partial\,\epsilon}|_{\epsilon=0}\right)\,y(\lambda+\epsilon)& \nonumber
\end{eqnarray}

\begin{eqnarray}
&a^{\dagger}\,y(\lambda)=\left( \left(\frac{4\,\rho\,Im \,z-r^2 }{4\,Re \,z}-i\,\rho \right)\,( \lambda^2-1)+ \frac{ i\, z}{2\,r }\,\lambda \right)  \,y(\lambda)&\\
&+\left( \left(\frac{4\,\rho\,Im \,z-r^2 }{4\,Re\, z}-i\,\rho \right)\,\frac{\partial^2}{\partial\,\epsilon^2}|_{\epsilon=0}-\left( \left(\frac{4\,\rho\,Im\, z-r^2 }{4\,Re\, z}-i\,\rho \right)\,2\,\lambda-\frac{ i\, z}{2\,r }\right)\,\frac{\partial}{\partial\,\epsilon}|_{\epsilon=0}\right)\,y(\lambda+\epsilon)& \nonumber
\end{eqnarray}

\begin{equation}
h\,y(\lambda)=i\,r\,\left( \frac{\partial}{\partial\,\epsilon}|_{\epsilon=0}\,y(\lambda+\epsilon)-\lambda\,y(\lambda)  \right)
\end{equation}

\noindent and

\begin{equation}
E\,y(\lambda)=y(\lambda)
\end{equation}

\medskip

\noindent (ii) If $z\in\mathbb{C}$ with $Re \,z=0$  then

\begin{eqnarray}
&a\,y(\lambda)=\left( \left(\rho-\frac{i\,Im \,z }{16\,r^2}\right)\,( \lambda^2-1)+r\,\lambda \right)  \,y(\lambda)&\\
&+\left( \left(\rho-\frac{i\,Im \,z }{16\,r^2} \right)\,\frac{\partial^2}{\partial\,\epsilon^2}|_{\epsilon=0}+\left( r-\left(\rho-\frac{i\,Im\, z }{16\,r^2} \right)\,2\,\lambda\right)\,\frac{\partial}{\partial\,\epsilon}|_{\epsilon=0}\right) \,y(\lambda+\epsilon) &\nonumber
\end{eqnarray}

\begin{eqnarray}
&a^{\dagger}\,y(\lambda)=\left( \left(\rho+\frac{i\,Im\, z }{16\,r^2}  \right)\,( \lambda^2-1)+r\,\lambda \right)  \,y(\lambda)&\\
&+\left( \left(\rho+\frac{i\,Im\, z }{16\,r^2} \right)\,\frac{\partial^2}{\partial\,\epsilon^2}|_{\epsilon=0}+\left(r- \left(\rho+\frac{i\,Im \,z }{16\,r^2} \right)\,2\,\lambda\right)\,\frac{\partial}{\partial\,\epsilon}|_{\epsilon=0}\right) \,y(\lambda+\epsilon) &\nonumber
\end{eqnarray}
   
\begin{equation}
h\,y(\lambda)=\frac{i\,Im\, z }{2\,r} \,\left( \frac{\partial}{\partial\,\epsilon}|_{\epsilon=0}\,y(\lambda+\epsilon) -\lambda\,y(\lambda) \right)
\end{equation}

\noindent and

\begin{equation}
E\,y(\lambda)=y(\lambda)
\end{equation}

\end{theorem} 

\begin{proof} The proof follows from theorem \ref{Schr} and (\ref{b1})-(\ref{b2}).

\end{proof}

\subsection{Random variables in $CEHeis$}

Self-adjoint operators $X$ on the Heisenberg Fock space
 $\mathcal{F}$ correspond to classical random variables with moment generating function $\langle \Phi , e^{s\,X}\,\Phi\rangle$ where $s\in\mathbb{R}$ and $\Phi$ is the Heisenberg Fock space cyclic vacuum vector such that $b\,\Phi=0$. In this subsection we compute the moment generating function of the self-adjoint operator $X=a+a^{\dagger}+h$.

\begin{lemma}\label{split}(Splitting formula) Let $L\in \mathbb{R}$ and $M, N\in\mathbb{C}$. Then for all $s\in\mathbb{R}$ such that $2\,L\,s+1>0$

\[
e^{s\,(L\,b^2 +L\,{b^{\dagger}}^2 -2\,L\,b^{\dagger}\,b -L+M\,b +N\,b^{\dagger})}\,\Phi=e^{w_1(s)\,{b^{\dagger}}^2}\,e^{w_2(s)\,b^{\dagger}}\,e^{w_3(s)}\,\Phi
\]

\noindent where

\begin{eqnarray}
w_1(s)&=&\frac{L\,s}{2\,L\,s+1}\label{DE1}  \\
w_2(s)&=&\frac{L\,(M+N)\,s^2+N\,s}{2\,L\,s+1}\label{DE2}    
\end{eqnarray}

\noindent and

\begin{equation}\label{DE3}
w_3(s)=\frac{(M+N)^2\,(L^2\,s^4+2\,L\,s^3)+3\,M\,N\,s^2  }{ 6\,( 2\,L\,s+1 ) }-\frac{\ln \, (2\,L\,s+1)}{2}
\end{equation}

\end{lemma} 

\begin{proof}  We will use  the differential method  of  proposition 4.1.1, chapter 1  of \cite{Fein}.  Let

\begin{eqnarray}\label{F}
F(s)&=&e^{s\,(L\,b^2 +L\,{b^{\dagger}}^2 -2\,L\,b^{\dagger}\,b -L+M\,b +N\,b^{\dagger})}\,\Phi\\
&=&e^{w_1(s)\,{b^{\dagger}}^2}\,e^{w_2(s)\,b^{\dagger}}\,e^{w_3(s)}\,\Phi\nonumber\\
\mbox{ (since $b^{\dagger}, {b^{\dagger}}^2$ and $ 1$  commute)  }&=&e^{w_1(s)\,{b^{\dagger}}^2+w_2(s)\,b^{\dagger}+w_3(s)}\,\Phi\nonumber
\end{eqnarray}

\noindent where $w_1, w_2, w_3$ are scalar-valued functions with $w_1(0)=w_2(0)=w_3(0)=0$. Then

\begin{equation}\label{F1}
\frac{\partial}{\partial\,s}\,F(s)=(w_1(s)\,{b^{\dagger}}^2+w_2(s)\,b^{\dagger}+w_3(s))\,F(s)
\end{equation}

\noindent and also

\begin{eqnarray}\label{F2}
\frac{\partial}{\partial\,s}\,F(s)&=&(L\,b^2 +L\,{b^{\dagger}}^2 -2\,L\,b^{\dagger}\,b -L+M\,b +N\,b^{\dagger})\,F(s)\\
&=&  (L\,b^2 +L\,{b^{\dagger}}^2 -2\,L\,b^{\dagger}\,b -L+M\,b +N\,b^{\dagger})\,e^{w_1(s)\,{b^{\dagger}}^2+w_2(s)\,b^{\dagger}+w_3(s)}\,\Phi\nonumber
\end{eqnarray}

\noindent Using lemma \ref{sf} (iii) with $f(b^{\dagger})=e^{w_1(s)\,{b^{\dagger}}^2+w_2(s)\,b^{\dagger}+w_3(s)} $  and the fact that $b\,\Phi=0$ we find that

\[
b\,F(s)=b\,f(b^{\dagger})\,\Phi=f^{\prime}(b^{\dagger})\,\Phi=(2\,w_1(s)\,b^{\dagger}+w_2(s))\,f(b^{\dagger})\,\Phi=(2\,w_1(s)\,b^{\dagger}+w_2(s))\,F(s)
\]

\noindent and

\begin{eqnarray*}
b^2\,F(s)&=&b\,(2\,w_1(s)\,b^{\dagger}+w_2(s))\,F(s)=(2\,w_1(s)\,(1+b^{\dagger}b)+w_2(s)\,b)\,F(s)\\
&=&( 2\,w_1(s)+ w_2(s)^2+4\,w_1(s)\,w_2(s)\,b^{\dagger}+4\,w_1(s)^2\,{b^{\dagger}}^2 )\,F(s)
\end{eqnarray*}

\noindent and so (\ref{F2}) becomes

\begin{eqnarray}\label{F3}
\frac{\partial}{\partial\,s}\,F(s)&=&\{2\,L\,w_1(s)+L\,w_2(s)^2-L+M\,w_2(s)\\
&&+(4\,L\,w_1(s)\,w_2(s)-2\,L\,w_2(s)+2\,M\,w_1(s)+N)\,b^{\dagger}\nonumber\\
&&+(4\,L\,w_1(s)^2+L-4\,L\,w_1(s))\,{b^{\dagger}}^2\}\,F(s)\nonumber
\end{eqnarray}

\noindent From (\ref{F1}) and  (\ref{F3}),  after equating coefficients of $1$, $b^{\dagger}$ and ${b^{\dagger}}^2$,  we  obtain
 
\begin{eqnarray*}
w_1^{\prime}(s)&=&4\,L\,w_1(s)^2-4\,L\,w_1(s)+L \mbox{  (Riccati differential equation)}\\
w_2^{\prime}(s)&=&(4\,L\,w_1(s)-2\,L)\,w_2(s)+2\,M\,w_1(s)+N  \mbox{  (Linear differential equation)}\\ 
w_3^{\prime}(s)&=&2\,L\,w_1(s)+L\,w_2(s)^2-L+M\,w_2(s)
\end{eqnarray*}

\noindent with $w_1(0)=w_2(0)=w_3(0)=0$. Therefore $w_1, w_2$ and $ w_3$ are given by (\ref{DE1})-(\ref{DE3}).

\end{proof}

\begin{remark}\end{remark} \noindent For $L\neq0$  the Riccati equation 

\[
w_1^{\prime}(s)=4\,L\,w_1(s)^2-4\,L\,w_1(s)+L
\]

\noindent appearing in the proof of lemma \ref{split} can be put in the canonical form

\[
V^{\prime}(s)=1+2\,\alpha\,V(s)+\beta\,V(s)^2
\]

\noindent of the theory of Bernoulli systems of chapters 5 and 6 of \cite{Fein}, where $V(s)=\frac{w_1(s)}{L}$, $\alpha=-2\,L$ and $\beta=4\,L^2$. Then $\delta^2:=\alpha^2-\beta=0$ which is characteristic of exponential and Gaussian systems (\cite{Fein},  Proposition 5.3.2). For $L=0$ we obtain classical Brownian motion (see proposition \ref{StPr} below).

\begin{proposition}\label{StPr}(Moment Generating Function) For all $s\in\mathbb{R}$  such that $2\,L\,s+1>0$

\begin{equation}\label{mgf}
\langle \Phi , e^{s\,(a+a^{\dagger}+h)}\,\Phi\rangle=(2\,L\,s+1)^{-1/2} \,e^{\frac{(M+N)^2\,(L^2\,s^4+2\,L\,s^3)+3\,M\,N\,s^2  }{ 6\,( 2\,L\,s+1 ) }} 
\end{equation}

\noindent where in the notation of theorem \ref{Schr}

\bigskip

\noindent (i) if $Re\, z\neq0$ then

\begin{eqnarray*}
L&=& \frac{4\,\rho\,Im \,z-r^2 }{2\,Re \,z}\\
M&=&-\left( \frac{Im \,z }{r}+i\,r\right)\\
N&=&-\left( \frac{Im \,z }{r}-i\,r\right)
\end{eqnarray*}

\bigskip

\noindent (ii) if $Re \,z=0$ then

\begin{eqnarray*}
L&=&2\,\rho\\
M&=&2\,r-i\,\frac{Im\, z}{2\,r}\\
N&=&2\,r+i\,\frac{Im\, z}{2\,r}
\end{eqnarray*}

\end{proposition}

\begin{proof} In both cases (i) and (ii) we find that

\[
a+a^{\dagger}+h=L\,b^2 +L\,{b^{\dagger}}^2 -2\,L\,b^{\dagger}\,b -L+M\,b +N\,b^{\dagger}
\]

\bigskip

\noindent Therefore, in the notation of  lemma \ref{split} using ${(e^{f(b^{\dagger})})}^*=e^{\bar f(b)}$ and the fact that for all scalars $\lambda$ we have that $e^{\lambda\,b}\,\Phi=\Phi$  we obtain

\begin{eqnarray*}
\langle \Phi , e^{s\,(a+a^{\dagger}+h)}\,\Phi\rangle&=&\langle \Phi , e^{s(L\,b^2 +L\,{b^{\dagger}}^2 -2\,L\,b^{\dagger}\,b -L+M\,b +N\,b^{\dagger}) } \,\Phi\rangle\\
&=&\,\langle \Phi ,e^{w_3(s) }\,\Phi\rangle\\
&=&(2\,L\,s+1)^{-1/2} \,e^{\frac{(M+N)^2\,(L^2\,s^4+2\,L\,s^3)+3\,M\,N\,s^2  }{ 6\,( 2\,L\,s+1 ) }}  \,\langle \Phi ,  \,\Phi\rangle\\
&=&(2\,L\,s+1)^{-1/2} \,e^{\frac{(M+N)^2\,(L^2\,s^4+2\,L\,s^3)+3\,M\,N\,s^2  }{ 6\,( 2\,L\,s+1 ) }}  
\end{eqnarray*}
\end{proof}

\begin{remark}\end{remark} \noindent If $L=0$ (corresponding to $\rho\,Im \,z > 0$ and $r^2=4\,\rho\,Im\, z$ in the case when $Re \,z\neq 0$ and to $\rho=0$ in the case when $Re \,z= 0$) then (\ref{mgf}) becomes 

\begin{equation}\label{BM}
\langle \Phi , e^{s\,(a+a^{\dagger}+h)}\,\Phi\rangle=e^{\frac{M\,N\,s^2  }{ 2} }=\begin{cases}
  e^{\left(\frac{(Im \,z)^2}{2\,r^2}+\frac{r^2}{2}  \right)\,s^2 } &\text{if $Re \,z\neq 0$ }\\
 e^{\left(2\,r^2+\frac{(Im \,z)^2}{8\,r^2} \right)\,s^2 }  &\text{if $Re \,z = 0$}
\end{cases}
\end{equation}

\noindent which means that $a+a^{\dagger}+h$ is a Gaussian random variable. 

\medskip

\noindent For  $L\neq0$ the term $(2\,L\,s+1)^{-1/2}$ appearing in (\ref{mgf}) is the moment generating function of a gamma random variable.

\subsection{Representation of $CEHeis$ in terms of two independent CCR copies}

 \begin{theorem}\label{CCR} For $j, k\in\{1,2\}$ let $\lbrack q_j, p_k \rbrack=\frac{i}{2}\,\delta_{j,k}$ and $\lbrack q_j, q_k \rbrack=\lbrack p_j, p_k \rbrack =0$ with $p_j^*=p_j$, $q_j^*=q_j$ and $i^2=-1$. 

\medskip

\noindent (i) If $z\in\mathbb{C}$ with $Re \,z\neq0$ and $Im \,z\neq0$ then  

\begin{eqnarray}
a&:=&i\,Re \,z\,q_1+\frac{1}{Re \,z}\,p_1^2-Im \,z\,p_2-\frac{i}{Im \,z}\,q_2^2\\
a^{\dagger}&:=&-i\,Re \,z\,q_1+\frac{1}{Re \,z}\,p_1^2-Im \,z\,p_2+\frac{i}{Im \,z}\,q_2^2 \\
h&:=&-2\,(p_1+q_2) 
\end{eqnarray}

\noindent and $E:=1$ satisfy the  commutation relations (\ref{extcommeq}) and the duality relations (\ref{1.1}) of $CEHeis$.

\medskip

\noindent (ii) If $z\in\mathbb{C}$ with $Re \,z=0$ and $Im \,z\neq0$ then for arbitrary $ r\in\mathbb{R}$ and $ c\in\mathbb{C}$ 

\begin{eqnarray}
a&:=&c\,p_1^2-Im \,z \,p_2+\left(r-\frac{i}{Im\,z}\right)\,q_2^2\\
a^{\dagger}&:=&\bar c\,p_1^2-Im \,z \,p_2+\left(r+\frac{i}{Im\,z}\right)\,q_2^2\\
h&:=& -2\,q_2
\end{eqnarray}

\noindent and $E:=1$ satisfy the  commutation relations (\ref{extcommeq}) and the duality relations (\ref{1.1}) of $CEHeis$.

\medskip

\noindent (iii) If $z\in\mathbb{C}$ with $Re \,z\neq0$ and $Im \,z=0$ then for arbitrary $ r\in\mathbb{R}$ and $ c\in\mathbb{C}$ 

\begin{eqnarray}
a&:=&i\,Re\,z\,q_1+\left(\frac{1}{Re\,z}+i\,r \right)\,p_1^2+c\,q_2^2\\
a^{\dagger}&:=&-i\,Re\,z\,q_1+\left(\frac{1}{Re\,z}-i\,r \right)\,p_1^2+\bar c\,q_2^2\\
h&:=&-2\,p_1
\end{eqnarray}

\noindent and $E:=1$ satisfy the  commutation relations (\ref{extcommeq}) and the duality relations (\ref{1.1}) of $CEHeis$.

\end{theorem} 

\begin{proof} (i) It is easy to see that $\lbrack q_j, p_j^2 \rbrack=i\,p_j$, $\lbrack q_j^2, p_j \rbrack=i\,q_j$ and $\lbrack q_j^2, p_j^2 \rbrack=2\,i\,p_j\,q_j$. Then

\begin{eqnarray*}
\lbrack  a, a^{\dagger} \rbrack&=&\frac{1}{Re \,z}\,i\,Re \,z\, \lbrack q_1 , p_1^2 \rbrack+\frac{1}{Re\, z}\,(-i\,Re \,z)\, \lbrack  p_1^2, q_1 \rbrack  -Im\, z \,\frac{i}{Im\, z}\,\lbrack p_2, q_2^2 \rbrack-\frac{i}{Im \,z}\,(-Im\, z) \,\lbrack q_2^2, p_2 \rbrack\\
&=&i\, (i\,p_1)-i\,(-i\,p_1)-i\,(-i\,q_2)+i\,(i\,q_2)\\
&=&-2\,(p_1+q_2) =h
\end{eqnarray*}

\begin{eqnarray*}
\lbrack  a, h \rbrack&=&\lbrack i\,Re \,z\,q_1+\frac{1}{Re \,z}\,p_1^2-Im \,z\,p_2-\frac{i}{Im \,z}\,q_2^2 , -2\,(p_1+q_2)\rbrack\\
&=& i\,Re \,z\,(-2) \lbrack  q_1, p_1 \rbrack-Im \,z\,(-2)\lbrack  p_2, q_2 \rbrack\\
&=& i\,Re \,z \,(-2)\,\left(\frac{i}{2}\right)-Im \,z\,(-2)\,\left(-\frac{i}{2} \right)\\
&=&Re \,z-i\,Im \,z=\bar z
\end{eqnarray*}

\noindent and

\begin{eqnarray*}
\lbrack  h, a^{\dagger} \rbrack&=&\lbrack-2\,(p_1+q_2), -i\,Re \,z\,q_1+\frac{1}{Re \,z}\,p_1^2-Im \,z\,p_2+\frac{i}{Im \,z}\,q_2^2    \rbrack\\
&=& 2\,i\,Re\, z \,\lbrack  p_1, q_1 \rbrack-2\,(-Im \,z)\,\lbrack  q_2, p_2 \rbrack\\
&=& 2\,i\,Re \,z \,\left(-\frac{i}{2}\right)+2\,Im \,z\, \frac{i}{2}\\
&=&Re \,z+i\,Im \,z  =z
\end{eqnarray*}

\noindent Clearly $(a^{\dagger} )^*=a$ and $h^*=h$. The proofs of (ii) and (iii) are similar.

\end{proof}

\begin{remark}\end{remark} \noindent In the notation of theorem \ref{CCR} we may take

\begin{equation}
q_1=\frac{b_1+b_1^{\dagger}}{2}\,\,\,;\,\,\,p_1=\frac{i\,(b_1^{\dagger}-b_1 )}{2}
\end{equation}

\noindent and 

\begin{equation}
q_2=\frac{b_2+b_2^{\dagger}}{2}\,\,\,;\,\,\,p_2=\frac{i\,(b_2^{\dagger}-b_2 )}{2}
\end{equation}

\noindent where 

\begin{equation}
\lbrack b_1, b_1^{\dagger} \rbrack=\lbrack b_2, b_2^{\dagger} \rbrack=1
\end{equation}

\noindent and

\begin{equation}
\lbrack b_1^{\dagger}, b_2^{\dagger} \rbrack=\lbrack b_1, b_2 \rbrack=\lbrack b_1, b_2^{\dagger} \rbrack= \lbrack b_1^{\dagger}, b_2\rbrack =0
\end{equation}

\noindent In that case theorem \ref{StPr} would extend to the product of the moment generating functions of two independent random variables defined in terms of the generators of two mutually commuting Schroedinger algebras. 

\section{The centrally extended Heisenberg group}

\begin{lemma}\label{l1}
For all $X, Y\in span\{ a, a^{\dagger}, h, E\}$
\[
e^{X+Y}=e^X\,e^Y\,e^{-\frac{1}{2}\,\lbrack X, Y\rbrack}\,e^{\frac{1}{6}\,\left(2\,\lbrack Y, \lbrack X, Y\rbrack \rbrack+ \lbrack X, \lbrack X, Y\rbrack \rbrack\right)}
\]

\end{lemma}

\begin{proof} This is a special case of the general Zassenhaus formula (converse of the BCH formula, see for example \cite{Su} and \cite{Fuchs}). In fact, using  (\ref{extcommeq})
we see that

\begin{equation}\label{comm-ce-ce2}
\lbrack CEHeis, CEHeis^{(1)} \rbrack
=\mathbb{C}\,E 
\end{equation} 

\noindent i.e. all triple commutators of elements of $span\{ a, a^{\dagger}, h, E\}$ are in the center.
\end{proof}

\begin{lemma}\label{l3} For all $\lambda, \mu \in \mathbb{C}$

\begin{equation}
e^{\lambda \, a}\,e^{\mu \, a^{\dagger}}=e^{\mu \, a^{\dagger} }\,e^{\lambda \, a}\,e^{\lambda\,\mu\,h}\,e^{\frac{\lambda\,\mu}{2}\, (\mu\, z-\lambda\,\bar z) }
\end{equation}

\begin{equation}
a\,e^{\mu \, a^{\dagger}}=e^{\mu \, a^{\dagger}}\,(a+\mu\,h+\frac{\mu^2\,z}{2})
\end{equation}

\end{lemma}

\begin{proof} By lemma \ref{l1} with $X=\mu \, a^{\dagger}$ and $Y=\lambda \, a $ we have

\begin{eqnarray*}
e^{\mu \, a^{\dagger}+\lambda \, a}&=&e^{\mu \, a^{\dagger}}\,e^{\lambda \, a}\,e^{-\frac{1}{2}\,\lbrack \mu \, a^{\dagger} , \lambda \, a\rbrack  }\,e^{\frac{1}{6}\,(2\,\lbrack \lambda \, a  ,   \lbrack \mu \, a^{\dagger} , \lambda \, a\rbrack  \rbrack     +\lbrack \mu \, a^{\dagger} ,   \lbrack \mu \, a^{\dagger} , \lambda \, a\rbrack  \rbrack)}\\
&=&e^{\mu \, a^{\dagger}}\,e^{\lambda \, a}\,e^{-\frac{\mu\,\lambda}{2}\,\lbrack  a^{\dagger} , a\rbrack  }\,e^{\frac{1}{6}\,(2\,\lambda^2\,\mu \,\lbrack  a  ,   \lbrack  a^{\dagger} ,  a\rbrack  \rbrack +\mu^2\,\lambda \,\lbrack  a^{\dagger} ,   \lbrack  a^{\dagger} ,  a\rbrack  \rbrack)}\\
&=&e^{\mu \, a^{\dagger}}\,e^{\lambda \, a}\,e^{\frac{\mu\,\lambda}{2}\,h}\,e^{\frac{1}{6}\,(-2\,\lambda^2\,\mu \,\bar z +\mu^2\,\lambda \,z)}
\end{eqnarray*}

\noindent Similarly, for $X=\lambda \, a $ and $Y=\mu \, a^{\dagger}$ we obtain

\[
e^{\lambda \, a+\mu \, a^{\dagger}}=e^{\lambda \, a}\,e^{\mu \, a^{\dagger}}\,e^{-\frac{\mu\,\lambda}{2}\,h}\,e^{\frac{1}{6}\,(-2\,\lambda\,\mu^2 \, z +\mu\,\lambda^2 \,\bar z)}
\]

\noindent Therefore

\[
e^{\mu \, a^{\dagger}}\,e^{\lambda \, a}\,e^{\frac{\mu\,\lambda}{2}\,h}\,e^{\frac{1}{6}\,(-2\,\lambda^2\,\mu \,\bar z +\mu^2\,\lambda \,z)}=e^{\lambda \, a}\,e^{\mu \, a^{\dagger}}\,e^{-\frac{\mu\,\lambda}{2}\,h}\,e^{\frac{1}{6}\,(-2\,\lambda\,\mu^2 \, z +\mu\,\lambda^2 \,\bar z)}
\]

\medskip

\noindent and so, multiplying both sides from the right by $e^{\frac{\mu\,\lambda}{2}\,h}\,e^{-\frac{1}{6}\,(-2\,\lambda\,\mu^2 \, z +\mu\,\lambda^2 \,\bar z)}$ we have that

\[
e^{\lambda \, a}\,e^{\mu \, a^{\dagger}}=e^{\mu \, a^{\dagger} }\,e^{\lambda \, a}\,e^{\lambda\,\mu\,h}\,e^{\frac{\lambda\,\mu}{2}\, (\mu\, z-\lambda\,\bar z) }
\]

\noindent which, after taking $\frac{\partial}{\partial\,\lambda}\,|_{\lambda=0}$ of both sides implies 

\[
a\,e^{\mu \, a^{\dagger}}=e^{\mu \, a^{\dagger}}\,\left(a+\mu\,h+\frac{\mu^2\,z}{2}\right)
\]
\end{proof}

\begin{lemma}\label{l2} Let $x$, $ D$  and $H$ be three operators satisfying the Heisenberg commutation relations

\begin{equation}
[D,x]_{Heis}=H, \,\,\,[D,H]_{Heis}=[x,H]_{Heis}=0
\end{equation}

\noindent Then, for all $s,b,c\in\mathbb{C}$

\begin{equation}
e^{s\,D}\,e^{b\,x}=e^{b\,x}\,e^{s\,D}\,e^{b\,s\,H}
\end{equation}

\end{lemma}

\begin{proof} The result is well known. A proof can be found in \cite{Fein}.
\end{proof} 

\begin{lemma}\label{l4}  For all $\lambda, \mu \in \mathbb{C}$

\begin{equation}
e^{\lambda \, a}\,e^{\mu \,h}=e^{\mu \,h}\,e^{\lambda \, a}\,e^{\lambda\,\mu\,\bar z}
\end{equation}

\begin{equation}
e^{\mu \,h}\,e^{\lambda \, a^{\dagger}}=e^{\lambda \, a^{\dagger}}\,e^{\mu \,h}\,e^{\lambda\,\mu\, z}
\end{equation}

\begin{equation}
a\,e^{\mu \,h}=e^{\mu \, h}\,(a+\mu\,\bar z)
\end{equation}

\begin{equation}
h\,e^{\lambda \, a^{\dagger}}=e^{\lambda \, a^{\dagger}}\,(h+\lambda\,z)
\end{equation}

\end{lemma}

\begin{proof} By commutation relations (\ref{extcomm}), $a$, $h$ and $\bar z$ are a copy of the Heisenberg algebra. Therefore, letting $D=a$, $x=h$ and $H=\bar z$ in lemma \ref{l2} we find that

\[
e^{\lambda \, a}\,e^{\mu \,h}=e^{\mu \,h}\,e^{\lambda \, a}\,e^{\lambda\,\mu\,\bar z}
\]

\noindent which, after taking adjoints and replacing $\bar \mu$ by $\mu$ and $\bar \lambda$ by $\lambda$, yields

\[
e^{\mu \,h}\,e^{\lambda \, a^{\dagger}}=e^{\lambda \, a^{\dagger}}\,e^{\mu \,h}\,e^{\lambda\,\mu\, z}
\]

\noindent and so  

\[
a\,e^{\mu \,h}=\frac{\partial}{\partial\,\lambda}\,|_{\lambda=0}\,e^{\lambda \, a}\,e^{\mu \,h}=\frac{\partial}{\partial\,\lambda}\,|_{\lambda=0}\,e^{\mu \,h}\,e^{\lambda \, a}\,e^{\lambda\,\mu\,\bar z}=e^{\mu \, h}\,(a+\mu\,\bar z)
\]

\noindent and 

\[
h\,e^{\lambda \, a^{\dagger}}=\frac{\partial}{\partial\,\mu}\,|_{\mu=0}\,e^{\mu \,h}\,e^{\lambda \, a^{\dagger}}=\frac{\partial}{\partial\,\mu}\,|_{\mu=0}\,e^{\lambda \, a^{\dagger}}\,e^{\mu \,h}\,e^{\lambda\,\mu\, z}   =e^{\lambda \, a^{\dagger}}\,(h+\lambda\,z)
\]

\end{proof}

\begin{corollary}\label{group}(Group Law) For $u, v, w ,y\in\mathbb{C}$ define

\begin{equation}\label{dfgrpel}
g( u, v, w,y):=e^{u \, a^{\dagger}} \, e^{v \,h}\,e^{w \, a}e^{yE}
\end{equation}

\noindent Then the family of operators of the form (\ref{dfgrpel}) is a group with group law given by

\begin{equation}\label{CEHeis-grplw}
g(\alpha, \beta, \gamma,\delta )\,g(A, B, C,D)=
\end{equation}

\[
=\, g(\alpha+A, \beta+B+\gamma\,A, \gamma+C,\left(\frac{\gamma\,A^2}{2}+ 
\beta\,A\right)\,z +\left(\frac{\gamma^2\,A}{2}+\gamma \,B \right)\,\bar z + \delta +D)
\]

\noindent  The family of operators of the form (\ref{dfgrpel}) with $u, v, w \in\mathbb{R}$ and $y\in\mathbb{C}$ is a sub--group. The group $\mathbb{R}^3\times \mathbb{C}$ endowed with the composition law:

\begin{equation}
(\alpha, \beta, \gamma,\delta)\,(A, B, C,D) =
\end{equation}

\[
\,\left(\alpha+A, \beta+B+\gamma\,A, \gamma+C,\left(\frac{\gamma\,A^2}{2}+ \beta\,A\right)\,z +\left(\frac{\gamma^2\,A}{2}+\gamma \,B \right)\,\bar z+\delta+D \right)
\]

\noindent is called the \textbf{centrally extended Heisenberg group}.
\end{corollary}

\begin{proof} Using lemmas \ref{l3} and \ref{l4} we have 

\begin{eqnarray*}
g(\alpha, \beta, \gamma,\delta)\,g(A, B, C,D)&=&e^{\alpha \, a^{\dagger}} \, e^{\beta \,h}\,e^{\gamma \, a}\,e^{A \, a^{\dagger}} \, e^{B \,h}\,e^{C \, a}e^{(\delta+DE)}\\
&=& e^{\alpha \, a^{\dagger}} \, e^{\beta \,h}\,e^{A \, a^{\dagger}} \,e^{\gamma\,a}\,e^{\gamma\,A\,h}\,e^{\frac{\gamma\,A}{2}\,(A\,z-\gamma\,\bar z)}\,e^{B \,h}\,e^{C \, a}e^{(\delta+D)E}\\
&=&e^{\alpha \, a^{\dagger}} \, e^{\beta \,h}\,e^{A \, a^{\dagger}} \,e^{\gamma\,a}\,e^{(\gamma\,A+B) \,h}\,e^{C \, a}e^{(\delta+D+\frac{\gamma\,A}{2}\,(A\,z-\gamma\,\bar z))E}\,\\
&=&e^{\alpha \, a^{\dagger}} \,e^{A \, a^{\dagger}} \, e^{\beta \,h}\,e^{\beta\,A\,z}\,e^{(\gamma\,A+B) \,h}\,e^{\gamma\,a}\,e^{ \gamma\,(\gamma\,A+B) \, \bar z }\,e^{C \, a}e^{(\delta+D+\frac{\gamma\,A}{2}\,(A\,z-\gamma\,\bar z))E}\\
&=&\,e^{(\alpha+A )\, a^{\dagger}} \, e^{(\beta+B+\gamma\,A)\,h}\,e^{(\gamma+C)\, a}
e^{\{\left(\frac{\gamma\,A^2}{2}+ \beta\,A\right)\,z +\left(\frac{\gamma^2\,A}{2}+\gamma \,B \right)\,\bar z+\delta+D\}E}
\end{eqnarray*}

\[
=\,g\left(\alpha+A, \beta+B+\gamma\,A, \gamma+C,\left(\frac{\gamma\,A^2}{2}+ \beta\,A\right)\,z +\left(\frac{\gamma^2\,A}{2}+\gamma \,B \right)\,\bar z+\delta+D \right)
\]

\end{proof}

\end{document}